\documentclass[aps,prb,preprint,a4paper,superscriptaddress,citeautoscript]{revtex4-1}
\usepackage[english]{babel}
\usepackage{lmodern}
\usepackage[T1]{fontenc}
\usepackage{amsmath}
\usepackage{units}
\usepackage{graphicx}
\usepackage{grffile}
\usepackage{color}
\usepackage[bottom]{footmisc}
\usepackage{booktabs}

\begin{document}

\title{Detailed optical spectroscopy of the hybridization gap and the hidden order transition in high quality URu$_{2}$Si$_{2}$ single crystals}

\author{N. Bachar}
\email[]{nimrod.bachar@unige.ch}
\affiliation{Department of Quantum Matter Physics, University of Geneva, Switzerland}

\author{D. Stricker}
\affiliation{Department of Quantum Matter Physics, University of Geneva, Switzerland}

\author{S. Muleady}
\affiliation{Department of Quantum Matter Physics, University of Geneva, Switzerland}

\author{K. Wang}
\affiliation{Department of Quantum Matter Physics, University of Geneva, Switzerland}

\author{J. A. Mydosh}
\affiliation{Kamerlingh Onnes Laboratory, Leiden University, 2300RA Leiden, The Netherlands}

\author{Y. K. Huang}
\affiliation{Van der Waals-Zeeman Institute, University of Amsterdam, 1018XE Amsterdam, The Netherlands}

\author{D. van der Marel}
\email[]{dirk.vandermarel@unige.ch}
\affiliation{Department of Quantum Matter Physics, University of Geneva, Switzerland}

\date{\today}

\begin{abstract}

We present a detailed temperature and frequency dependence of the optical conductivity measured on clean high quality single crystals of URu$_{2}$Si$_{2}$ of $ac$- and $ab$-plane surfaces. Our data demonstrate the itinerant character of the narrow 5f bands, becoming progressively coherent as temperature is lowered below a cross-over temperature $T^*{\sim}75~K$. $T^*$ is higher than in previous reports as a result of a different sample preparation, which minimizes residual strain. We furthermore present the density-response (energy-loss) function of this compound, and determine the energies of the heavy fermion plasmons with $a$-and $c$-axis polarization. Our observation of a suppression of optical conductivity below 50~meV both along $a$ and $c$-axis, along with a heavy fermion plasmon at 18~meV, points toward the emergence of a band of coherent charge carriers crossing the Fermi energy and the emergence of a hybridization gap on part of the Fermi surface. The evolution towards coherent itinerant states is accelerated below the hidden order temperature $T_{HO}=17.5$~K. In the hidden order phase the low frequency optical conductivity shows a single gap at $\sim 6.5$meV, which closes at $T_{HO}$.
\end{abstract}


  
\maketitle

\section{Introduction}\label{sec:intro}

The intriguing properties of the heavy fermion compound URu$_{2}$Si$_{2}$~\cite{Palstra1985,Maple1986} and its unique temperature-pressure-field phase diagram~\cite{Kim2003,Villaume2008,Hassinger2008} constitute a source of inspiration for unconventional states of matter, the debates about which have all but abated~\cite{Mydosh2011,Mydosh2014}. 
URu$_{2}$Si$_{2}$ is a compensated metal; the bands close to the Fermi energy have predominantly uranium $5f_{5/2}$ character with $j_z=\pm 5/2$ and $j_z=\pm 3/2$~\cite{Oppeneer2010,Oppeneer2011}. 
Apart from superconductivity observed below $T_c=1.5$ K~\cite{Maple1986,Schlabitz1986}, a relatively low superconducting transition temperature common for heavy fermions materials, the states of matter at higher temperatures defy the understanding of the underlying magnetic and electronic properties of URu$_{2}$Si$_{2}$. 

Below $T_{HO}=17.5$~K a hidden-order (HO) is observed as a partial gap opening in both spin and charge excitation spectra~\cite{Palstra1985,Maple1986,Bonn1988,Thieme1995,Degiorgi1997,Hall2012,Nagel2012,Guo2012,Lobo2015,Buhot2014,Kung2015,Schmidt2010,Aynajian2010,Hasselbach1992,Thieme1995,Rodrigo1997,Park2012}. Recently, high resolution synchrotron X-ray diffraction measurements at zero field revealed a small orthorhombic symmetry breaking distortion at $T_{HO}$~\cite{Tonegawa2014}. Haule and Kotliar \cite{Haule2009} argued that the U-sites should be in an entangled state of two $5f^2$ configurations with azimuthal spin-orbital momentum $J=4$ and opposite magnetic quantum numbers $J_z=\pm 4$, corresponding to a hexadecapolar state carrying zero angular momentum. Recent polarization resolved Raman data were found to be consistent with hexadecapolar nature of primary A$_{2g}$ character, rendered chiral by admixture of A$_{1g}$ character~\cite{Kung2015}. Inter-site ordering of these local configurations leads to a commensurate chirality density wave~\cite{Kung2015}. Consistency with torque data~\cite{Okazaki2011} would require an additional admixture of a B$_{1g}$ component~\cite{Kung2015}.

Above $T_{HO}$ no long range order of any kind is observed, but spectroscopic experiments indicate a partial gapping that disappears gradually for temperatures higher than $T^*\sim75$~K (the coherence temperature).
In a typical heavy fermion compound, hybridization between narrow (in energy) $d$ or $f$ bands with broad  $sp$ bands leads to a metallic state with renormalized properties attributed to the coherent electrons~\cite{Dordevic2001}. Indeed, the DC resistivity $\rho(T)$ reveals a crossover at $T^*$ from a high temperature phase where $d\rho/dT <0$ to a phase with $d\rho/dT >0$  at low temperatures~\cite{Palstra1986}. Since the uranium $5f$ states are on the border between localized and itinerant, the natural interpretation of the cross-over is, that itinerant heavy electron bands emerge at low temperature due to hybridization of the uranium $5f$ states with the $spd$ electrons of the ligands~\cite{Mydosh2011}. The crossover at  $T^*$ and its impact on the electronic properties is still controversial~\cite{Nagel2012}; many interpretations insist that a hybridization gap appears only at $T_{HO}$. Optical conductivity spectra are particularly effective for analyzing the electrodynamic properties associated to various states of electrically conducting matter, such as Fermi liquid, superconductor, the hidden order phase of URu$_2$Si$_2$ and the hybridization gap state emerging below $T^*$. Here we use optical spectroscopy to probe the detailed temperature and frequency dependence of the optical conductivity in URu$_{2}$Si$_{2}$ single crystals.  We demonstrate full agreement with photo-emission, quantum oscillation, scanning tunneling and point contact spectroscopy data. Our optical work reinforces contact and stimulates comparison to electronic structure calculations based on density functional theory\cite{Oppeneer2010,Oppeneer2011} and dynamical mean field theory\cite{Haule2009}. Previous optical conductivity measurements~\cite{Levallois2011,Guo2012} and point contact spectroscopy measurements~\cite{Park2012} showed that a hybridization gap emerges in the coherent state well above $T_{HO}$, at temperatures of the order of 35~K while its impact on the $c$-axis polarization is still not clear~\cite{Lobo2015}. In the present paper we revisit this effect and show that after ultra high vacuum annealing of the crystals to relax strain, the cross-over occurs at ${\sim}75~K$. We determine the heavy fermion plasmon energy from investigation of the energy loss function, demonstrating the emergence of coherent itinerant heavy charge carriers below $T^{*}$. If indeed strain in the crystals effectively shifts the cross-over to a lower temperature, this would explain the spread of values of the cross-over temperature reported with different surface conditions and different experimental techniques. In addition, we observe that the HO phase is characterized by a single gap of 6.5~meV.

\par 

The organization of this paper is as follows:
First, we review our experimental methods (Section~\ref{sec:expr}) used to obtain the polarization dependent reflectivity data (Section~\ref{sec:mainres}). We then show that the optical conductivity exhibits a gradual reduction in both $a$-axis and $c$-axis as the temperature is lowered below $T^*$ (Section~\ref{sec:conc}) which can be attributed to new inter-band transitions forming as a result of a modified electronic energy bands at the coherent state (Section~\ref{sec:hybr}). The value of $T^*$ in the present work is higher than in previous optical experiments \cite{Levallois2011} owing to a different sample preparation, minimizing strain. We also address the conductivity spectral weight transfer due to the coherent and hidden order states (Section~\ref{sec:hybr}) and the Fano line-shape of the phonon modes due to coupling to the electronic continuum (Section~\ref{sec:fano}). In section~\ref{sec:sigmaHO} we show that a high-quality strain free surface exhibits the well-known HO gap feature in the $a$-axis and $c$-axis optical conductivity spectra with a single gap structure. Conclusions are drawn in section~\ref{sec:conc}.           

\section{Experimental}\label{sec:expr}

In this work, we studied large single crystals of URu$_{2}$Si$_{2}$ grown by means of the traveling zone method in the mirror furnace in Amsterdam under high purity (6N) argon atmosphere. The zone melting is believed to be helpful for keeping the impurity concentration low in the crystal. Indeed, neutron scattering measurements confirm the smallest reported value of~$m_{s}~{\approx}~0.012\mu_{B}$ for the anti-ferromagnetic moment per uranium atom~\cite{Amitsuka2007,Niklowitz2010}. The oriented crystals were cut using the spark erosion method to create samples with $ac$-plane and $ab$-plane crystal surfaces. X-ray Laue diffraction was used to verify the orientation of the crystal surfaces, and to orient the $a$ and $c$ axes of the $ac$-plane surface for optical experiments. 

\par

One $ac$-plane sample and two $ab$-plane samples from two different crystal batches of URu$_{2}$Si$_{2}$ were polished. The samples were then annealed for 48 hours at 950$^{\circ}$C in a ultra high vacuum chamber using the procedure described recently by Buhot \textit{et al.}~\cite{Buhot2015}. The annealing procedure aims at relaxing residual strain introduced by polishing. For the $ac$-plane surface this procedure successfully produced mirror-like surfaces with the URu$_{2}$Si$_{2}$ stoichiometry. In the following sections we will present the results obtained on these strain- and oxygen-free ac-surfaces. Applying the same procedure to the $ab$-plane surface did not result in stoichiometric URu$_{2}$Si$_{2}$ and in particular, unlike the $ac$-plane, showed the presence of UO$_2$ on the surface. 
  
\par

\begin{figure}
\begin{center}
\includegraphics[width=0.5\linewidth]{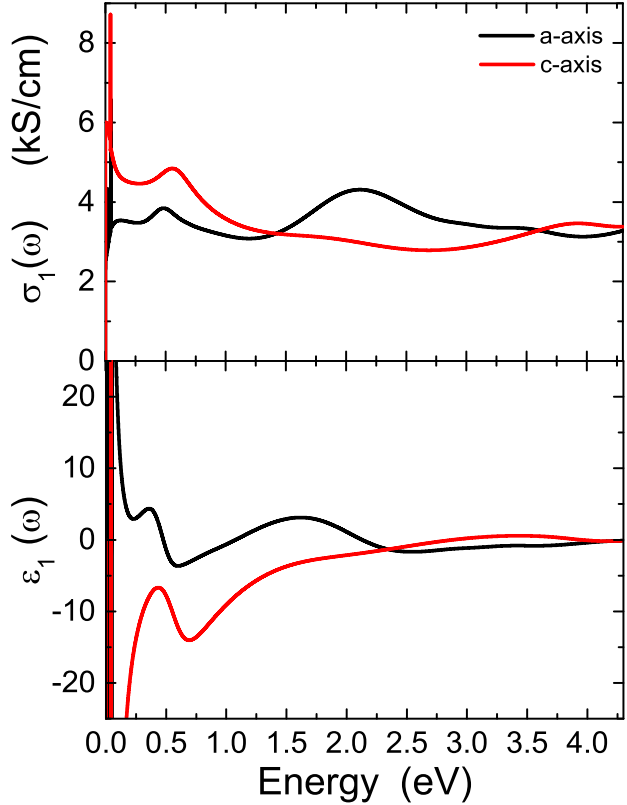}
\caption{\label{fig:sigma-loss-RT}
(Color online) Room temperature (290 K) optical conductivity (top) and dielectric function (bottom) of URu$_{2}$Si$_{2}$ with electric field polarized along the $a$-axis (black) and $c$-axis (red). These data were obtained on the $ac$-plane surface of a URu$_{2}$Si$_{2}$ single crystal combining spectroscopic ellipsometry along two orthogonal planes of reflection.}
\end{center}    
\end{figure}

The optical response of the aforementioned crystals was measured by combining infrared reflectivity and ellipsometry. In the photon energy range of 2.5~meV to 0.45~eV measurements were obtained using near normal incident reflectivity using a Fourier transform infrared spectrometer. Detailed temperature dependence was recorded in the infrared regime from room temperature down to 9~K by using a ultra high vacuum helium flow cryostat. The cryostat is designed to maintain a high position stability of the mounted sample, thus enabling a continuous measurement as the sample is cooled down to low temperatures. As a result we obtain approximately 1~K temperature resolution in our data. Calibration spectra were obtained by measuring the gold layer deposited on top of the single crystal using in-situ thermal evaporation. Polarizers for the different frequency regimes were used in order to measure separately the reflectivity of the \textbf{a} and $c$-axis in the $ac$-plane single crystal. The crystal was aligned by Laue X-ray diffraction to ensure that the $c$-axis was in vertical orientation therefore minimizing admixture of the $a$-axis in the optical measurements. In the photon energy range of 0.45~eV to 4.4 eV the complex dielectric function was determined using a Woollam VASE\textsuperscript{\textregistered} spectroscopic ellipsometer. For this end the ratios of the reflection coefficients for $p$ and $s$ polarized light $\rho=r_p/r_s$, were measured at incident angles of 70$^{\circ}$ and 73$^{\circ}$ with the reflection plane along the $a$-axis and the $c$-axis. In order to fully disentangle  $a$-axis and  $c$-axis contributions the four data sets (two polarizations and two angles of incidence) were fitted to a Drude-Lorentz model (See appendix~\ref{app:ellips}). Combining the fit of the ellipsometry measurements with the reflection measurements at lower frequencies resulted in the dielectric functions shown in Figure~\ref{fig:sigma-loss-RT} for the range of the experimental data. The ellipsometry data were measured at room temperature while the reflectivity below 0.5 eV was measured between 9 and 300 K in 1 K steps. The optical conductivity below 0.5 eV was obtained by Kramers-Kronig analysis, using the reflectivity from 0.5 eV  to $\infty$ calculated from aforementioned Drude-Lorentz fits along $a$-axis and $c$-axis. Here we relied on the assumption that the spectra above 0.5 eV have weak temperature dependence, which is motivated by observation of very weak temperature dependence of the optical reflectivity in the mid-infrared range. 
This procedure anchors the phase output of the Kramers-Kronig transformation of the reflectivity data in the entire frequency range~\cite{Mirzaei2013}. For extrapolating the lower frequencies we have compared two methods: (i) a wide Drude peak was fitted to the reflectivity data at temperatures above the HO phase and a narrow Drude peak which was fitted to the low frequency part of the data where a gap-like feature is observed, (ii) the Hagen-Rubens extrapolation. These two approaches resulted in negligible differences of the real and imaginary parts of the optical conductivity in the frequency range of the actual data. 

\section{Results and Discussion}

\subsection{Main features of the experimental data}\label{sec:mainres}

The real part of the optical conductivity and dielectric function are presented in Figure~\ref{fig:sigma-loss-RT} for the full measured frequency range, and for electric field polarized along the $a$ and the $c$ direction. Peaks in $\sigma_1{(\omega)}$ correspond to transverse polarized excitations, including optical phonons (sharp features below 0.1 eV) and interband transitions weighted by the optical matrix elements. Longitudinal excitations, in particular plasmons, are revealed by the zero-crossing of the dielectric function $\epsilon(\omega)$. The small peak in the optical conductitivity at $\sim 0.5$ eV (both polarizations) and the larger one at $\sim 2$ eV ($a$-axis) confirm the optical conductivity spectra reported by Degiorgi obtained from Kramers-Kronig analysis of broad-band reflectivity data of a polished $ab$-plane surface\cite{Degiorgi1997}, but with a higher intensity in particular for the  $\sim 2$~eV peak, possibly due to differences in surface condition.  

\begin{figure}
\begin{center}
\includegraphics[width=0.5\linewidth]{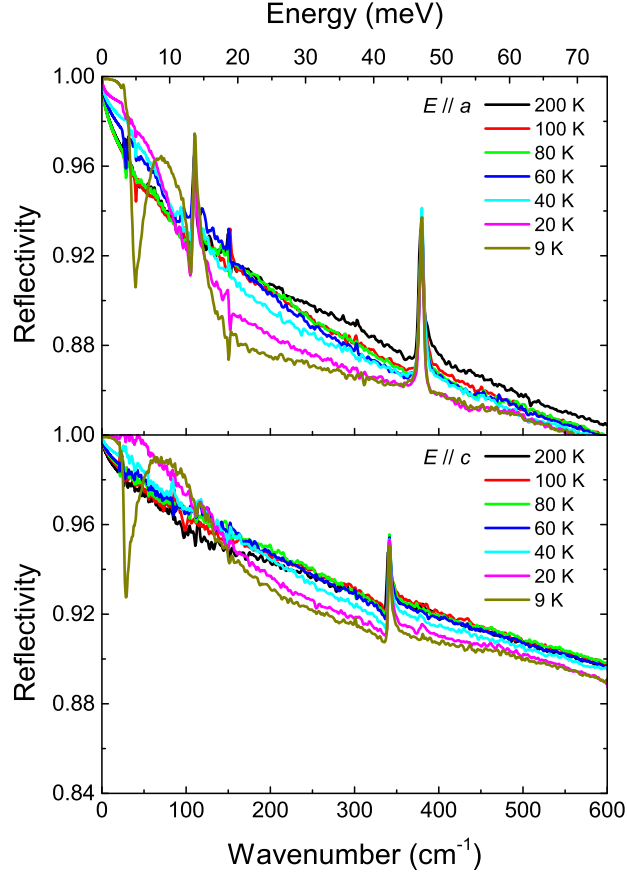}
\caption{\label{fig:ref-ac}
(Color online) Reflectivity of URu$_{2}$Si$_{2}$ single crystal in $a$-axis (upper panel) and $c$-axis (lower panel) polarizations for selected temperatures.}
\end{center}
\end{figure}

The reflectivity in the FIR (far infrared) frequency range is shown in Figure~\ref{fig:ref-ac} for selected temperatures. The well known anisotropy of URu$_{2}$Si$_{2}$ can be observed in Fig.~\ref{fig:sigma-loss-RT} where the $c$-axis data shows higher reflectivity than the $a$-axis data as it is more metallic. The $a$-axis data (Fig.~\ref{fig:ref-ac}) shows two phonons at $\hbar\omega_{0}{\approx}13.5$~meV (108.8~cm$^{-1}$) and $\hbar\omega_{0}{\approx}46.9$~meV (378.5~cm$^{-1}$) corresponding to the E$_{u}$(1) and E$_{u}$(2) modes, respectively~\cite{Buhot2015}. Glitches at 150 cm$^{-1}$ (Fig.~\ref{fig:ref-ac}a) and 90  cm$^{-1}$ (Fig.~\ref{fig:ref-ac}b) are due to instrument noise (150 cm$^{-1}$) and merging of two spectral ranges (90  cm$^{-1}$). The temperature evolution of the $a$-axis reflectivity data in the FIR range can be tracked by looking at two different frequency regimes. Below 10~meV the reflectivity increases dramatically for temperatures lower than 80~K consistent with the known coherence temperature $T^*\sim 75~K$ of URu$_{2}$Si$_{2}$. At lower temperatures, \textit{ i.e.} below $T_{HO}$, the reflectivity develops a dip at approximately 5~meV with a relative decrease of about 7\% (given by the value of R$_{20K}$/R$_{9K}$ at  5~meV) which is higher than previous reported values~\cite{Levallois2011,Hall2012,Lobo2015}. At frequencies above and below aforementioned HO feature, the reflectivity increases as a function of decreasing temperature, \textit{e.g.} for the measured data at approximately 10~meV and for the extrapolated data at 0.25~meV due to a narrowing Drude peak. Above $\sim$10~meV, the reflectivity develops a non-monotonic decrease as a function of frequency. The decrease in the reflectivity is clearly observed at frequencies around 20~meV and develops rapidly below $T^*$ as will be further illustrated by the optical conductivity curves.        

In the FIR freqeuncy range the $c$-axis reflectivity (Fig.~\ref{fig:ref-ac}) is higher than that of the $a$-axis by about 6\% (given \textit{e.g.} R$_{c}$/R$_{a}$ at $\omega/2{\pi}c=600~cm^{-1}$), in agreement with the known anisotropy of URu$_{2}$Si$_{2}$ already reported in previous works~\cite{Levallois2011} showing that the $c$-axis response is more metallic. The $c$-axis data reveals two phonons at $\hbar\omega_{0}{=}14.1$~meV (114.7~cm$^{-1}$) and $\hbar\omega_{0}{=}42.1$~meV (340~cm$^{-1}$) corresponding to the A$_{2u}$(1) and A$_{2u}$(2) modes, respectively~\cite{Buhot2015}. The low energy mode is observed only at low temperatures in our measurements. Similar to the $a$-axis data, the reflectivity frequency dependence is altered below $T^*$. A slight increase of the reflectivity is observed below 12~meV which is enhanced toward the HO phase. At higher frequencies, approximately 25~meV, the reflectivity decreases with a non-monotonic frequency dependence; this temperature dependence is less pronounced than in the $a$-axis data. The HO optics signature in the form of a reflectivity dip evolves also in the $c$-axis direction as shown in Fig.~\ref{fig:ref-ac}. 

\begin{figure}[ht!]
\begin{center}
\includegraphics[width=0.5\linewidth]{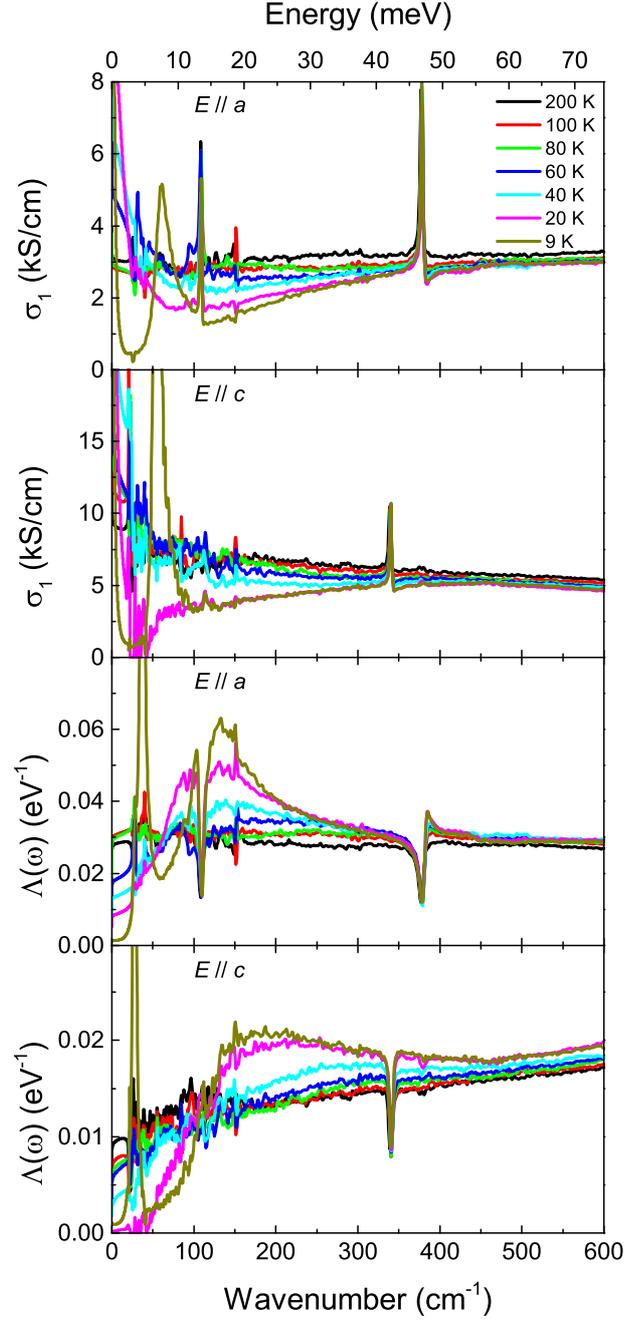}
\caption{\label{fig:siglosPG-ac}
(Color online) Optical conductivity (top) and energy loss function (bottom) of URu$_{2}$Si$_{2}$ single crystal in $a$-axis and $c$-axis polarizations for selected temperatures.}
\end{center}
\end{figure}

\subsection{Temperature dependence of the optical conductivity through $T^*$}\label{sec:sigma}

In Figure~\ref{fig:siglosPG-ac} the real part of conductivity, $\sigma_{1}$, is displayed in the FIR frequency regime for the $a$-axis and $c$-axis (top panels). In the lower two panels the quantity  
\begin{equation}\label{eq:loss}
 \Lambda(\omega)=\frac{2}{\pi\hbar\omega}\mbox{Im}\frac{-1}{\epsilon(\omega)}
\end{equation} 
is displayed for the $a$-axis and $c$-axis. $\Lambda(\omega)$ is directly proportional to the energy loss function multiplied by a factor $1/\omega$. The energy loss-function corresponds to the charge response function, and peaks at frequencies corresponding to longitudinal modes, in particular at the free carrier plasmon resonance (to be discussed in section~\ref{sec:hybr}). 
At high temperatures, the $a$-axis conductivity shows almost frequency independent values of about 3~kS/cm except for the two phonons at 108.8~cm$^{-1}$ and 378.5~cm$^{-1}$. Below $T^*$, \textit{e.g.} T~=~60~K, we observe a non-monotonic suppression of $\sigma_{1}$ for frequencies of approximately 10 to 50~meV while the low frequency data show a dramatic increase. Along the c-axis $\sigma_{1}(\omega)$ has already at 200~K a broad Drude peak, which is superimposed on a frequency-independent background of 5~kS/cm. Below $T^*$ it shows a gradual conductivity increase at low frequencies, while below 40~K the change is much more dramatic and suppression of the FIR conductivity can be clearly observed in our data.           
Levallois \textit{ et al.} \cite{Levallois2011} observed below a cross-over temperature of about 30~K a suppression of the optical conductivity in a broad range up to 50~meV with minima around 15~meV for the $a$-axis and 11~meV for the $c$-axis. The relatively low value of aforementioned cross-over temperature as compared to $T^*{\approx}75~K$ observed in transport experiments, was probably the result of strain in the samples. Based on the argument that no amplitude fluctuations of the hidden order gap are observed in the specific heat in this temperature range, Levallois \textit{ et al.} \cite{Levallois2011} proposed that the partial gap below $T^*$ is a hybridization gap, and also concluded that the corresponding electronic configuration could be a precursor of the hidden order state. Guo \textit{ et al.} \cite{Guo2012} reported a rapid suppression of the $a$-axis optical conductivity below 50 K, with a minimum at 17~meV associated with the development of a Drude component at low temperature, and an almost full recovery of the spectral weight near 600~meV. Based on the complete absence of the HO feature at 8~meV above the HO temperature, Guo \textit{ et al.} argued that the hybridization gap can not be taken as a precursor of the HO state \cite{Guo2012}.  Indeed, while the coherent state sets the stage allowing the hidden-order to stabilize at $T_{HO}$ and as such is likely a prerequisite to the HO state, this does not imply -nor does it require- the presence of pre-transitional fluctuations of the hidden-order parameter.

\begin{figure}
\begin{center}
\includegraphics[width=0.5\linewidth]{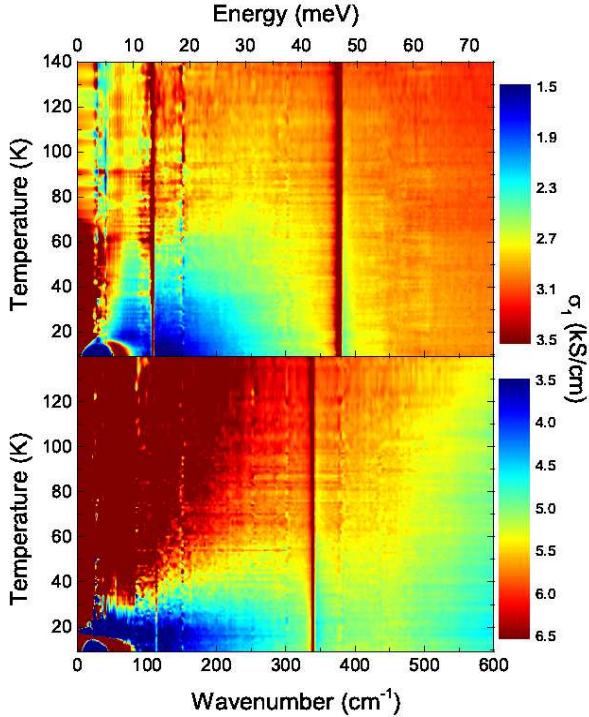}
\caption{\label{fig:sig-ac-2D}
(Color online) Detailed colormap of the optical conductivity as a function of temperature and frequency for $a$-axis (upper panel) and $c$-axis (lower panel) polarizations. In the $a$-axis, the optical conductivity becomes suppressed below $T^*$ in the form of dome-like shape while in the $c$-axis direction, it is observed as a narrowing Drude peak as temperature is lowered below $T^*$. The HO gap appears as a dome-like shape suppression of the conductivity in both axes.}
\end{center}
\end{figure}

Our experimental method allows to measure the optical spectra with a temperature resolution of 1~K. Figure~\ref{fig:sig-ac-2D} shows a color map of the conductivity as a function of temperature and frequency. The aforementioned suppression of the $a$-axis conductivity in the FIR range is observed as a dome-like shape starting to form below $T^*$ and centered at about 18~meV. Concomitantly with the conductivity suppression, a sharp increase in $\sigma_{1}$ is observed at low frequencies, \textit{ i.e.} below 5~meV. An additional conductivity suppression is observed below $T_{HO}$ forming a second dome-like shape at frequency of approximately 4~meV, which is bordered by a narrow Drude peak and piled-up conductivity at a frequency of approximately 7~meV.      

\par        

The $c$-axis conductivity shown in Figure~\ref{fig:sig-ac-2D} has a significant Drude contribution already at high temperatures, slightly narrowing as temperature is lowered down to $T^*$. Below that temperature the Drude peak narrows more rapidly, accompanied by a suppression of $\sigma_{1}$ in the same frequency range as for the $a$-axis data. Additional suppression of $\sigma_{1}$ can be seen below $T_{HO}$ as a dome-like shape centered at approximately 3~meV. Similar to the $a$-axis data, the HO feature is also bordered by a narrow Drude peak and pile up of conductivity at a frequency of approximately 7~meV.

\par 

\begin{figure}
\begin{center}
\includegraphics[width=0.5\linewidth]{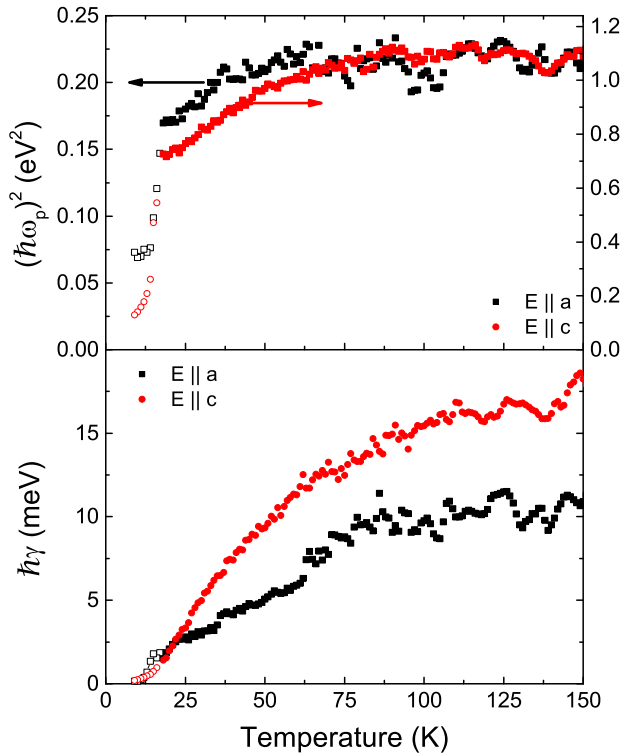}
\caption{\label{fig:drude_param}
(Color online) The spectral weight $\omega_{p}^2$ (upper panel) and the scattering rate $\gamma$ (lower panel) of the Drude contribution to the optical conductivity obtained from fitting our reflectivity data to the Drude-Lorentz model (Full symbols). $\omega_{p}^2$ decreases below $T^*$ with almost same temperature dependence for both $a$-axis and $c$-axis, however with higher values for the $c$-axis. $\gamma$ is drastically reduced at low temperatures and shows a clear change of slope as a function of temperature in the $a$-axis data compared to the $c$-axis data. Below $T_{HO}$ the Drude peak becomes too narrow for the reflectivity data, that begin at 2.5 meV. The parameters in this temperature range (open symbols) were obtained by fitting the low frequency part (below 4 meV) of $\sigma_1(\omega)$ (Fig. \ref{fig:siglosPG-ac}) instead of fitting the reflectivity data.}
\end{center}
\end{figure}

In order to address the coherent part of the conductivity ({\em i.e.} the narrow zero-frequency mode) we use a Drude-Lorentz fit to our data, and concentrate on the parameters of the Drude peak, namely its spectral weight $\omega_{p}^{2}$ and scattering rate $\gamma$, shown in Figure~\ref{fig:drude_param} as a function of temperature. The spectral weight is fairly unchanged at high temperatures while it reduces below $T^*$ from 0.22~eV$^2$ to 0.17~eV$^2$ ($a$-axis) and 1.05~eV$^2$ to 0.75~eV$^2$ ($c$-axis) for the $a$-axis and $c$-axis respectively. Assuming a constant density of carriers we can estimate that the increase of the effective mass below $T^*$ and down to $T_{HO}$ is of the order of 30\% and 40\% for $a$-axis and $c$-axis respectively, which is consistent with previous reports showing that the mass renormalization in the coherent state is lower than expected from a typical heavy fermion compound~\cite{Stewart1984}. At the same temperature, the scattering rate is reduced drastically reflecting the reduced width of the Drude peak, also observed in Figure~\ref{fig:siglosPG-ac} as the appearance of the dark red area at low frequencies in the $a$-axis data and in the $c$-axis data.
 
\begin{figure}
\begin{center}
\includegraphics[width=0.5\linewidth]{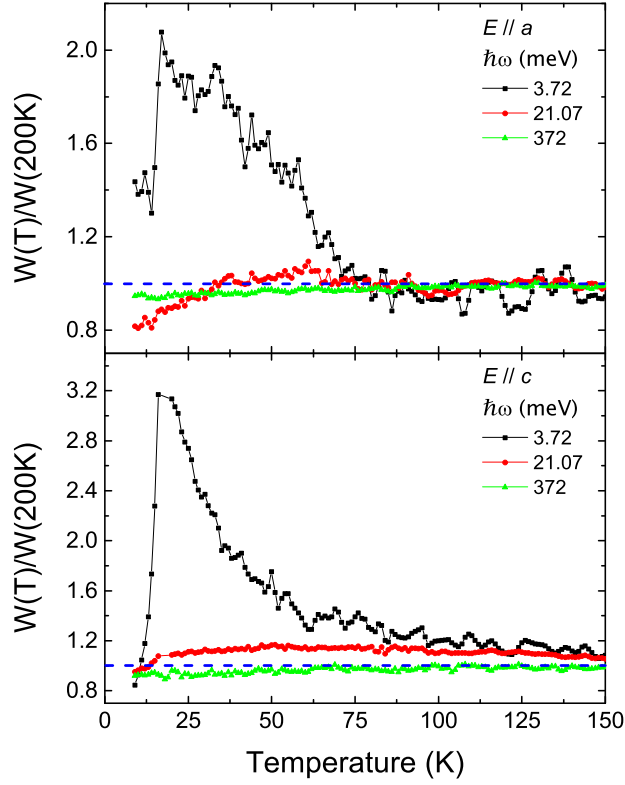}
\caption{\label{fig:delta_sw}
(Color online) normalized spectral weight obtained by integrating the conductivity from zero to the cut-off frequency $\omega$ indicated in the figure legend for the $a$-axis (upper panel) and $c$-axis (lower panel) data. The spectral weights are normalized to W(200~K).}
\end{center}
\end{figure}

As shown in Figure~\ref{fig:sig-ac-2D}, the conductivity suppression observed at approximately 20~meV and 4~meV below $T^*$ and $T_{HO}$, respectively, is concomitant with an enhancement of the conductivity at low frequencies, \textit{ i.e.} the Drude peak, and pile up conductivity just above the HO gap. To analyze the temperature dependence in further detail we calculate the normalized spectral weight
\begin{equation}
W(\omega,T) = \int_{0}^{\omega} \sigma_{1}({\omega}') d{\omega}'
\end{equation}
where $\omega$ is the upper frequency for the accumulated spectral weight. The temperature dependence of $W(\omega,T) $ is displayed in Figure~\ref{fig:delta_sw} for three cut-off frequencies. Taking the cutoff at $\sim 4$~meV we observe a clear increase of $W(4~\mbox{meV},T)$ below $T^*$, which is caused by the narrowing of the free carrier zero-frequency mode, consistent with the analysis of the Drude parameters (see Fig. \ref{fig:drude_param}). This behavior suggests a change of the state of matter of the free charge carriers. Cooling further, the slope of $W(4~\mbox{meV},T)$  exhibits a break at $T_{HO}$. This temperature dependence of the low energy spectral weight is similar to that of the second order phase transition in {\em e.g.}  a superconductor.For the $a$-axis ($c$-axis) $W(4~\mbox{meV},9K)$ drops to approximately 70\% (25\%) of its value at $T_{HO}$.  The spectral weight lost at low frequencies due to the HO transition is recovered by the conductivity piled up just above the gap (at approximately 7~meV) as shown by $W(21~\mbox{meV},9K)$. Along the $c$-axis (but not along the $a$-axis) we observe that $W(21~\mbox{meV},9K)$ has a small drop below $T_{HO}$, suggesting that less than 4\% of the free carrier spectral weight is transferred to frequencies above 21~meV, confirming the recent results of Lobo {\em et al.}~\cite{Lobo2015}.

\subsection{Hybridization and inter-band transitions}\label{sec:hybr}

The physics of heavy fermion compounds is often described as hybridization phenomena involving electrons in broad itinerant bands and electrons in narrow bands of strongly interacting quasi-atomic states, \textit{e.g.} the $5f$ electrons in URu$_{2}$Si$_{2}$. At sufficiently low temperatures  itinerant metallic bands may emerge as a result, where the conduction electrons' effective mass is renormalized as can for example be observed as an enhancement of the electronic specific heat. Such a crossover to renormalized metallic behavior affects the optical conductivity as well~\cite{Basov2011}.
The coherent part of the free carrier response, represented by the narrow zero-frequency mode in the optical conductivity, is indeed changing drastically as a function of temperature and frequency below $T^*$ as shown in Figures~\ref{fig:siglosPG-ac} and~\ref{fig:sig-ac-2D}. In the range of 5 to 50~meV the $a$-axis and $c$-axis conductivities reveal a suppression of the optical conductivity. The evolution of these spectral features through $T^*$ occurs very gradually, as expected for a crossover phenomenon.     

Another way of examining the low energy charge dynamics is by analyzing $\Lambda(\omega)$ defined in Eq. \ref{eq:loss}. When integrated over the full frequency range this quantity satisfies the sum-rule~\cite{Mahan2000}
\begin{equation}
 \int_0^{\infty}\Lambda(\omega)d\omega=1
\end{equation} 
At low frequencies $\Lambda(\omega)$ is proportional to $\tau^{-1}(\omega)/\omega_{p}^2$, where $\tau^{-1}(\omega)$ is the optical scattering rate in the so-called extended Drude analysis~\cite{Goetze1972}. However, when the frequency ranges of intra-band and interband transitions overlap, the determination of the $\omega_{p}^2$ factor becomes ambiguous. Since the latter ambiguity is actually present in the case of URu$_2$Si$_2$, we will focus here on $\Lambda(\omega)$.

\par

The two lower panels of Figure~\ref{fig:siglosPG-ac} give $\Lambda(\omega)$ as obtained from the $a$-axis and $c$-axis data. $\Lambda(\omega)$ shows a drastic change as temperature is lowered below $T^*$. The curves above $T^*$ show an almost frequency independent value while below $T^*$ a peak emerges at a frequency of approximately 18~meV. At \textit{e.g.} 60~K, $\Lambda(\omega)$ starts to develop some frequency dependence. In the $a$-axis data a distinct peak shows up in  $\Lambda(\omega)$. Along the $c$-axis the low temperature $\Lambda(\omega)$ is enhanced in a broad range from 15 to 50~meV, but there is no peak. 
If no damping were to be present, the peak occurs exactly at the frequency where $\epsilon(\omega)$ crosses zero, which in the present case corresponds to the screened plasma-resonance of the heavy charge carriers that form below $T^*$.  Comparing the peak-frequency of 18~meV with the fitted Drude parameter in Fig. \ref{fig:drude_param}, we note that the former is almost a factor 20 smaller than the latter. This difference is caused by the screening of the free carrier plasmon by the interband transitions. The peak at $\sim 18$~meV  is broadened due to the overlap with the region of inter-band transitions in $\sigma_1(\omega)$, causing it to be over-damped above $T^*$, and progressively less damped when temperature drops below  $T^*$.  
The fact that the peak only shows up below $T^*$ is therefor a consequence of the simultaneous emergence of a narrow zero-frequency mode due to the renormalized free charge carriers, and a gap-like suppression of optical conductivity in the far-infrared range. 
Note that, while optical conductivity~\cite{Levallois2011,Guo2012}, ultrafast optics~\cite{Liu2011}, point contact spectroscopy~\cite{Park2012,Park2014} and angle resolved photo-emission spectroscopy~\cite{Boariu2013} data have indicated a temperature dependent cross-over well above the HO transition where low-energy spectral weight is removed, values for the cross-over temperature vary from 30~K to 60~K, possibly due to different surface strain conditions.   
 
\begin{figure}
\begin{center}
\includegraphics[width=0.5\linewidth]{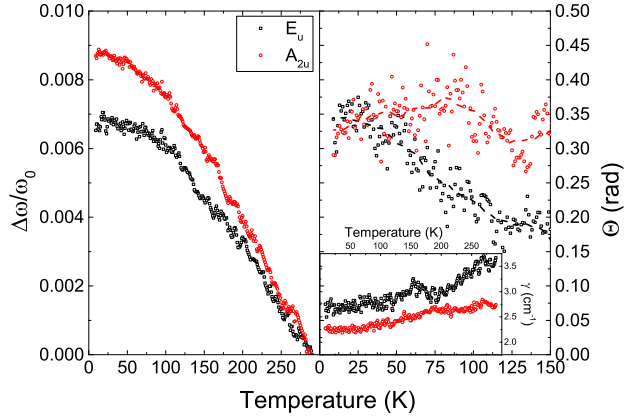}
\caption{\label{fig:FanoParam} 
(Color online) Summary of the three fit parameters of the high energy phonon modes of the $a$-axis and $c$-axis to a Fano-shape resonance where $\Delta\omega=\omega(T)-\omega(0)$ and $\omega_{0}$ is the oscillator's central frequency at room temperature.}
\end{center}
\end{figure} 
 
The temperature evolution of the electronic excitations of URu$_{2}$Si$_{2}$ can be understood as follows. The main electronic degrees of freedom that we need to consider are a partly filled narrow band of primarily $5f$ character in which the interactions are strong, and a partly filled wide band of more itinerant character where the interactions are weak. At sufficiently high temperatures coherent charge transport of the narrow band is wiped out by thermal fluctuations. The narrow-band electrons behave as double degenerate localized $5f^2$ multi-electron states. $\Lambda(\omega)$ then shows a smooth frequency dependence associated to the inelastically scattered wide-band electrons. Decreasing temperature through $T^*$ restores coherence of the narrow band transport, and at the same time the avoided crossing at the Fermi energy due to hybridization of the bands becomes noticeable. This behavior is described by the Kondo-lattice model, and it is responsible for the formation of a band of coherently conducting free charge carriers with strong mass renormalization. 
Returning to the optical conductivity of the paramagnetic phase at low temperatures but still above $T_{HO}$, we expect that interband transitions occur within the manifold of renormalized heavy fermion bands. These interband transitions are present for all frequencies starting at $\omega=0$, due to the presence of contours in momentum space where bands intersect at $E_F$. This is illustrated in Fig. 16 of Ref. \onlinecite{Oppeneer2010} for certain trajectories between $\Gamma-M$ and $\Gamma-X$. (Note that, even if the many-body renormalization of the single-particle spectral function is not included in the band calculations, quantum oscillation experiments~\cite{Hassinger2010} and photo emission~\cite{Santander-Syro2009,Yoshida2010,Boariu2013,Meng2013,Chatterjee2013} suggest that the prediction of the Fermi-surface is nevertheless correct.) Our interpretation of the optical conductivity spectrum below 50~meV and its temperature dependence down to $T_{HO}$ is therefore, that at low temperatures (but still above $T_{HO}$) the  gradually increasing conductivity with increasing frequency is due to interband-transitions within the manifold of $5f_{5/2}$ bands crossing the Fermi-energy. These bands partly overlap with each other and accommodate optical transitions from zero frequency upwards. At high enough temperatures (on the scale of $T^*$) the inverse lifetime of the electrons becomes large, and the $5f$ electrons loose their itinerant character. This inverses the temperature trend of the DC resistivity. Since the mean-free path becomes very short, the optical excitations involve wave-packages on the order of the interatomic distance rather than itinerant Bloch-states, resulting in the broad and featureless optical conductivity observed above $T^*$.
 
\subsection{Phonon Fano-line shape}\label{sec:fano}

The conductivity color map (Figure~\ref{fig:sig-ac-2D}) reflects also a temperature evolution of the E$_{u}$ and A$_{2u}$ phonon modes in the $a$-axis and $c$-axis, respectively. Below $T{\approx}40K$ our spectral reveal the A$_{2u}$(1) phonon mode of the $c$-axis; this feature is less pronounced than in recent reports~\cite{Buhot2015,Lobo2015}. The high frequency E$_{u}$ and A$_{2u}$ phonon modes exhibit an asymmetric Fano shape resonance which is gradually evolving below $T^*$. In order to obtain the oscillator's center frequency $\omega_{0}$, width $\gamma$, oscillator strength $S$ and asymmetry angle $\Theta$ we have fitted our data to the expression 
\begin{equation}
\sigma(\omega) = 
\frac{S\omega_0^2}{4\pi}\frac{\omega+i\omega_0\tan\Theta}{\gamma\omega+i(\omega_{0}^{2}-\omega_{}^{2})}
\end{equation} 
Figure~\ref{fig:FanoParam} shows that the higher frequency E$_{u}$ and A$_{2u}$ modes undergo a gradual change as temperature is lowered. The central frequency $\omega_{0}$ increases gradually and tends to saturate at lower temperatures with approximately 0.7 to 0.9\% increase for the $a$-axis and $c$-axis, respectively. The non-zero values of $\Theta$ indicates significant coupling of the the E$_{u}$ and A$_{2u}$ phonon modes to the continuum of electronic excitations in this energy range. The asymmetry of the E$_{u}$ phonon diminishes almost a factor two when temperature is raised to 150 K. The A$_{2u}$ asymmetry on the other hand has only weak temperature dependence. Furthermore, the Fano line-shape of these optical phonons is neither affected in a significant way by the emergence of coherent transport, nor by the hidden order. 

\begin{figure}[ht!]
\begin{center}
\includegraphics[width=0.5\linewidth]{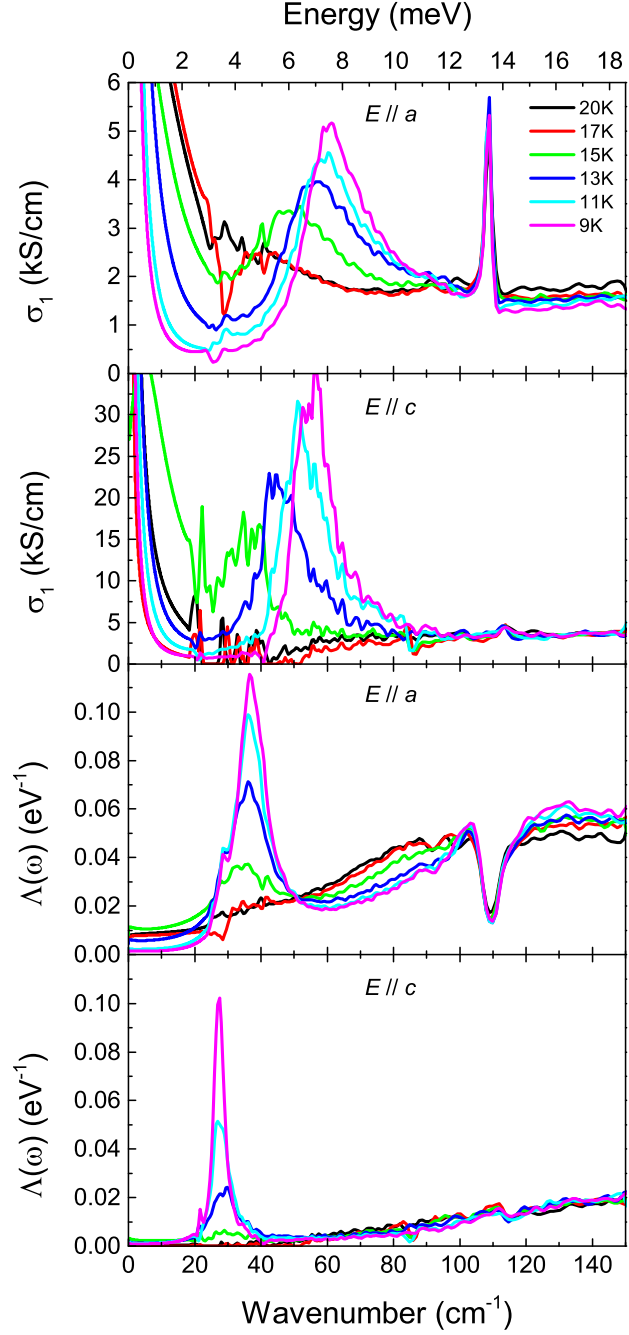}
\caption{\label{fig:sig-HO} 
(Color online) Low energy optical conductivity (top panels) and loss function (bottom panels) at selected temperatures close and below $T_{HO}$ for $a$-axis and $c$-axis polarizations. }
\end{center}
\end{figure}
    
\subsection{Low energy charge response in the Hidden Order phase}\label{sec:sigmaHO}
The detailed temperature dependence of the conductivity in the hidden order phase and the corresponding loss functions are shown in Figure~\ref{fig:sig-HO}. In both axes, a narrow Drude peak with scattering rate, $\hbar\tau^{-1}$, of the order of 0.1~meV is forming while a pile up of conductivity is shown at higher frequency of approximately 7~meV. A conductivity suppression in $\sigma_{1}$ below 5~meV evolves gradually with decreasing temperature. The midpoint of the conductivity rise on the left side of the peak is almost identical for the two polarizations (6.3 and 6.5 meV for the $a$-axis and the $c$-axis respectively) implying weak or non-existant anisotropy in the HO optical gap. Along with the opening of the HO gap we observe the emergence of a sharp longitudinal mode along both axes, with energies 4.6 and 3.5 meV respectively. These peaks reveal an anisotropic plasmon mode of the renormalized charge carriers, corresponding to the reduced carrier density in the HO phase. Given the collective nature of these modes their energies and anisotropy have no one-to-one relation to the single-particle gap.  A large anisotropy is also visible in the conductivity along the different crystal orientations, where $\sigma_{1}$ peaks at about 5~kS/cm and 30~kS/cm for the $a$-axis and $c$-axis respectively. This is also born out by the change of the free carrier spectral weight, shown in Fig.~\ref{fig:drude_param}, indicating a drop by a factor of 2.5 and 7 for the $a$-axis and $c$-axis respectively. The observation of a transition into a phase with fewer charge carriers confirms the prediction of Oppeneer~\textit{et al.}~\cite{Oppeneer2010} of a strong Fermi surface reconstruction upon entering the low temperature ordered phase. 
  
\begin{figure}
\begin{center}
\includegraphics[width=0.5\linewidth]{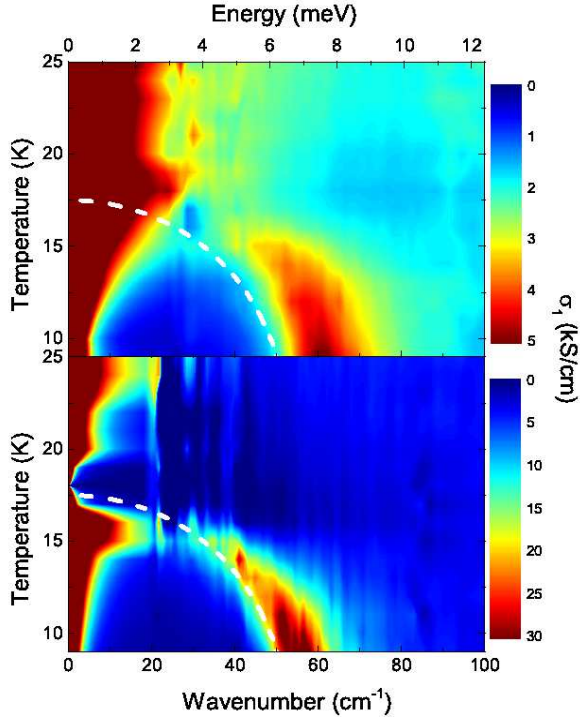}
\caption{\label{fig:sig2D-HO}
(Color online) Detailed temperature and frequency dependence of the optical conductivity at the HO phase transition. The gap closes and follows a BCS gap-like temperature dependence with $T_{HO}$ of 17.5~K and $2\Delta_{HO}=6.5~meV$ (Dashed white line) for both axes (upper panel $a$-axis; lower panel $c$-axis.}
\end{center}
\end{figure}

The values of the conductivity peak of $\sigma_{1}$ for both axes can be compared to previous works reporting a gap-like feature in the HO phase. For the $a$-axis data we find only small but still evident difference between value of approximately 4~kS/cm~\cite{Hall2012} to the value of 5~kS/cm   reported for our crystals and others~\cite{Lobo2015}. For the $c$-axis we find a value ranging from approximately 4.5~kS/cm~\cite{Hall2012} to approximately 15~kS/cm~\cite{Lobo2015} while our reported value is much higher, \textit{i.e.} 30~kS/cm at $T=10$~K. 
In Figure~\ref{fig:sig2D-HO} we show a zoom into the HO gap relevant temperature and frequency regimes of $\sigma_{1}$ measured for both axes. We see, that the temperature dependence of the gap is similar for both axes, described by a BCS-like temperature dependence. Given $T_{HO}=~17.5~K$ and $2\Delta_{HO}~=~6.5~meV$ we arrive at $2\Delta_{HO}/k_BT_{HO}=4.3$, which is slightly larger than the BCS value $3.5$. 

\par
\begin{figure}
\begin{center}
\includegraphics[width=0.5\linewidth]{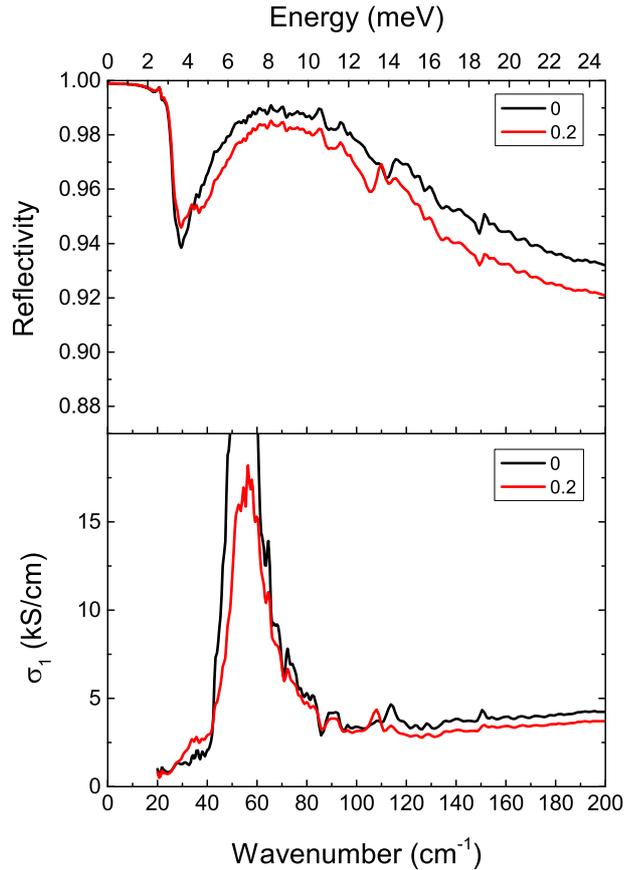}
\caption{\label{fig:sig-mix}
(Color online) Reflectivity (upper panel) and optical conductivity (lower panel) simulations of an $a$-axis and $c$-axis admixture. The 20\% $a$-axis inclusions depicted in our simulations by a polarization leakage is showing a spurious multiple-gap structure which results in reduced conductivity and an additional low frequency slope change compared to that of the clean $c$-axis data.}	
\end{center}
\end{figure}

In contrast to previous publications~\cite{Hall2012,Lobo2015}  we do not observe a multiple gap component in the $c$-axis data. We attribute these discrepancies to differences in crystal quality and/or homogeneity:
Figure~\ref{fig:sig-mix} shows a simulation obtained by mixing the $a$-axis and $c$-axis polarization dependent reflectivity. We compare the reflectivity and conductivity of our $c$-axis data to the simulated data taking into account a leakage of 20\% of the $a$-axis data. The reflectivity of the mixed axes reveals a multiple gap structure similar to that of Ref.~\onlinecite{Hall2012} having two minima in the HO phase at frequency of approximately 5~meV. The conductivity of the $c$-axis having an initially higher value than the one reported by Hall \textit{et al.}~\cite{Hall2012} is reduced to values similar to those reported by Lobo \textit{et al.}~\cite{Lobo2015}. In addition, the low frequency part shows a bump-like feature which is clearly higher than the clean $c$-axis data, while a feature in the form of an additional change of slope appears in the conductivity of the mixed axes simulated data at approximately 6~meV. We note that $c$-axis leakage can occur for a variety of reasons, only some of which are faithfully modeled by the simulation in Fig. \ref{fig:sig-mix}. The two minima in the reflectivity at 4.6 and 3.5~meV match the longitudinal plasma frequencies displayed in Fig. \ref{fig:sig2D-HO}; these features propagate to the optical conductivity and are responsible for the spurious conductivity around 4~meV (30 cm$^{-1}$).  Comparison to the conductivity curves reported by Hall \textit{et al.}~\cite{Hall2012} and Lobo \textit{et al.}~\cite{Lobo2015}, leads us to postulate that the reported multiple gap feature may be due to admixture of $a$-axis and $c$-axis contributions. 

The presence of a partial gap of several meV in the HO phase has been previously reported based on specific heat and transport data~\cite{Maple1986,Palstra1986}, and low energy optical, Raman, tunneling, and point contact spectra~\cite{Bonn1988,Thieme1995,Degiorgi1997,Hall2012,Nagel2012,Guo2012,Lobo2015,Buhot2014,Kung2015,Schmidt2010,Aynajian2010,Hasselbach1992,Thieme1995,Rodrigo1997,Park2012}. 
Initially the hidden order was interpreted as a magnetic order~\cite{Maple1986,Broholm1987}. However, the moments in the range 0.01-0.03 $\mu_{B}$ vary strongly from one sample to another~\cite{Matsuda2001,Amitsuka2002,Yokoyama2005,Amitsuka2007} and are too small to account for the large entropy change at the HO transition~\cite{Broholm1987,Broholm1991}. Above $T_{HO}$ inelastic neutron scattering~\cite{Wiebe2007,Villaume2008,Janik2009,Bourdarot2010} reveals a magnetic mode having its minimal energy of 4.8~meV at wavevector $Q_{1}=(1\pm0.4,0,0)$. Below $T_{HO}$ an additional mode emerges having its minimum energy of 1.7~meV at wavevector $Q_{0}=(1,0,0)$. Based on analysis of quantum oscillations in the hidden order phase 5 different Fermi pockets were identified with effective masses ranging from $2.5$ to $30$ $m_{e}$\cite{Hassinger2010}. Quasi-particle interference imaging by scanning tunneling microscopy shows splitting of a light conduction band into two heavy fermion bands only below $T_{HO}$~\cite{Schmidt2010}. Photo-emission~\cite{Santander-Syro2009,Yoshida2010,Boariu2013,Meng2013,Chatterjee2013} and Shubnikov-de Haas measurements~\cite{Hassinger2010} showed a Fermi surface reconstruction upon entering the HO state with square-like shape and additional half circle Fermi petals~\cite{Bareille2014}. It was also demonstrated that the different experimental observables can be correlated, \textit{e.g.} the response of the $A_{2g}$ Raman mode and the $Q_{0}$ wave vector~\cite{Buhot2014} or alternatively the Fermi surface $k$-space geometry with the $Q_{1}$ wave vector below $T_{HO}$~\cite{Bareille2014}, thus suggesting the spin and charge degrees of freedom to be strongly coupled in URu$_{2}$Si$_{2}$. Taken together these observations suggest that the hidden transition is accompanied by the reconstruction of narrow itinerant bands crossing the Fermi energy. The folding and shifting of the bands below $T_{HO}$ changes the number of itinerant charge carriers. The electron-hole compensation inherent to this compound allows in principle the carrier concentration to cancel out entirely, but a finite but small amount of electrons and holes remain in the HO phase, and transitions between the folded bands show up as a distinct gap in low energy spectroscopic probes. Since aforementioned low energy modes observed with neutron and Raman scattering are of collective nature, their energies don't necessarily match any interband transition. Since the gap is set by the band dispersion, it shouldn't depend on polarization of the electromagnetic field in the infrared experiments, and for this reason we observe the same gap value (6.5~meV) along $a$- and $c$ axis. Differences of the optical spectra above the gap have to do with differences in optical matrix elements connecting occupied and empty bands, as well as the dispersion thereof in momentum space.

\section{Conclusions}\label{sec:conc}

We have observed the cross-over of the optical conductivity of URu$_{2}$Si$_{2}$ from the high temperature phase to a low temperature phase, still well above $T_{HO}$, characterized by a suppression of optical conductivity exhiting a minimum at 17~meV, confirming previous reports of a partial gapping far above $T_{HO}$\cite{Levallois2011,Guo2012}. The present data on strain-free surfaces indicate the cross-over to take place at around $T^*\sim 75$K, which is a higher temperature than in the previous data on differently prepared surfaces, observed along both the $a$ and $c$-axis.
The emergence of the hybridization gap is accompanied by a gradual increase of the effective mass of the free charge carriers, as revealed by a progressive reduction of the Drude spectral weight upon cooling, and a a maximum in the energy loss function at 18~meV, providing a direct signature for the emergence of a plasmon of the heavy charge carriers. This provides clear evidence for the itinerant character existing already in the hybridization gap phase below $T^*$, which continues to evolve at an accelerated rate when temperature drops below $T_{HO}$. Our data in the HO phase show a single HO gap feature. Our findings confirm by and large the behavior anticipated on the basis of density functional theory and dynamical mean field theory\cite{Oppeneer2010,Haule2009}.

\begin{acknowledgments}

We thank A. V. Balatsky, G. Blumberg, K. Haule, R.P.S.M. Lobo, A. B. Kuzmenko, J. Levallois, J. Teyssier and P. de Visser for useful discussions. This work was supported by the Swiss National Science Foundation (SNSF) through Grants 200021-153405 and 200021-162628.
 
\end{acknowledgments}

\begin{figure}[ht!]
\begin{center}
\includegraphics[width=0.5\linewidth]{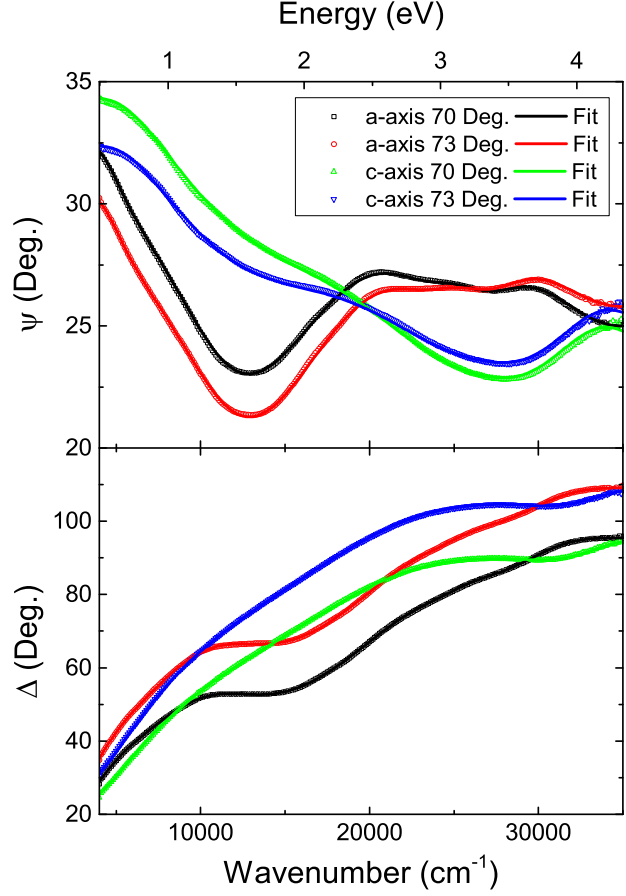}
\caption{\label{fig:psi_delta_ac}
(Color online) $\Psi(\omega)$ and $\Delta(\omega)$ spectra at 300~K for two planes of reflection of the $ac$-plane of URu$_{2}$Si$_{2}$ measured at two angles of incidence. Experimental data are shown together with a simultaneous fit of all eight curves to the Drude-Lorentz parametrization of the $a$-axis and $c$-axis dielectric functions. Experimental data and fitted curves can not be distinguished because they fully overlap.}
\end{center}
\end{figure}
\begin{figure}[ht!]
\begin{center}
\includegraphics[width=0.5\linewidth]{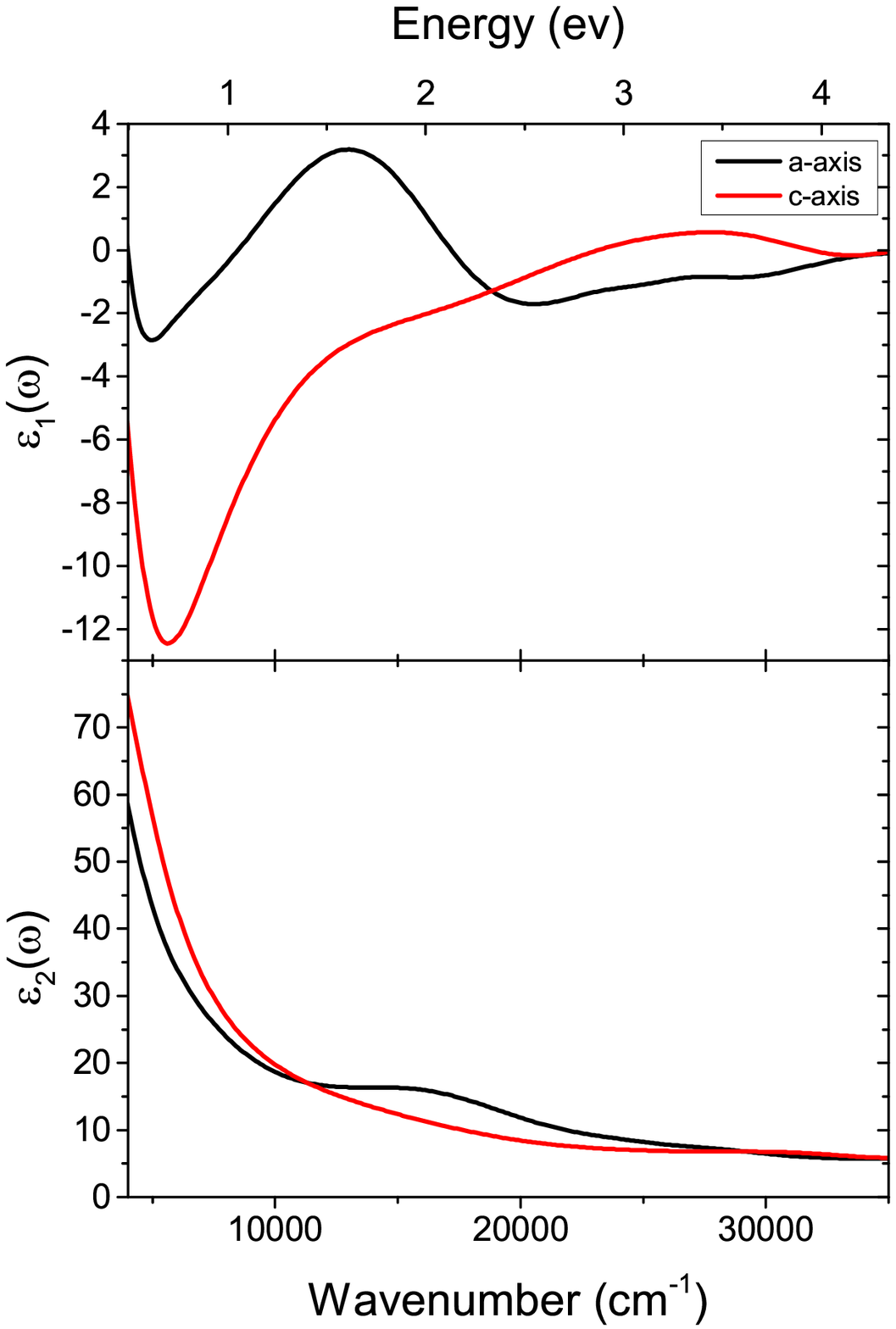}
\caption{\label{fig:sigma_loss_fit}
(Color online) The real (upper panel) and imaginary (lower panal) parts of the dielectric function $\epsilon(\omega)$ $a$-axis (black) and $c$-axis (red) polarizations obtained from a fit to the Drude-Lorentz model and based on the ellipsometry measurements shown in Fig.~\ref{fig:psi_delta_ac}.}
\end{center}
\end{figure}
\appendix*
\section{Ellipsometry measurements of URu$_{2}$Si$_{2}$}\label{app:ellips}
We measured the complex dielectric function using a Woollam VASE\textsuperscript{\textregistered} spectroscopic ellipsometer in the energy range from 0.48~eV to 4.3~eV. The reflectivity ratio for $p$ and $s$ polarization is defined as
\begin{equation}
	r_{p}/r_{s} = \tan(\Psi)e^{i\Delta}
	\nonumber
\end{equation}
where $\Psi$ and $\Delta$ are the parametric amplitude and phase difference components. 
For a crystal with optical axis along $x$, $y$ and $z$ and a surface along the $xy$-plane the dependence of the $r_{p}/r_{s}$ ratio on the tensor elements of the dielectric function is described by the Fresnel equation
\begin{equation}
\frac{r_{p}}{r_{s}}= \frac{\sqrt{\epsilon_x}\cos{\theta}-\sqrt{1-\sin^2\theta/\epsilon_z}} {\sqrt{\epsilon_x}\cos{\theta}+\sqrt{1-\sin^2\theta/\epsilon_z}} \cdot \frac{\cos{\theta}+\sqrt{\epsilon_y-\sin^2\theta}}{\cos{\theta}-\sqrt{\epsilon_y-\sin^2\theta}}
\nonumber	
\end{equation}
In the present case the surface of the tetragonal crystal is along the $ac$-plane, so that $z=a$.  We measured the $r_{p}/r_{s}$ ratios at incident angles of 70$^{\circ}$ and 73$^{\circ}$ with the reflection plane intersecting with the $a$-axis ($x=a$, $y=c$) and the $c$-axis of the crystal surface ($x=c$, $y=a$). The corresponding $\Psi(\omega)$ and $\Delta(\omega)$ spectra are displayed in Figure~\ref{fig:psi_delta_ac}. We fitted all  $\Psi(\omega)$ and $\Delta(\omega)$ spectra (two angles of incidence for each of the two planes of reflection) simultaneously using a Drude-Lorentz parametrization of $\epsilon_a(\omega)$ and $\epsilon_c(\omega)$. 
The good quality of these fits, displayed in Figure~\ref{fig:psi_delta_ac}, is manifested by the complete overlap of experimental data and fitted curves. The corresponding spectra of the real and imaginary parts of $\epsilon_a(\omega)$ and $\epsilon_c(\omega)$ are displayed in Figure~\ref{fig:sigma_loss_fit}. Together with the reflectivity results we calculated the full range spectra displayed in Figure~\ref{fig:sigma-loss-RT} of the main text.  

\newpage
\newpage
\bibliographystyle{apsrev4-1}
\bibliography{PRB2016_Bachar_URu2Si2}
\end{document}